\documentclass{article}
\usepackage[preprint]{spconfa4}
\usepackage{amsmath,graphicx,url,times}
\usepackage{color}
\usepackage{xcolor,colortbl}
\usepackage[mode=buildnew]{standalone}
\usepackage{multirow}
\usepackage{booktabs}
\usepackage{etoolbox, siunitx}
\usepackage{tabularx}
\usepackage{xcolor}
\usepackage{tikz}
\usepackage{hyperref}
\usepackage{pgfplots} 

\usetikzlibrary{dsp}
\usetikzlibrary{chains}
\usetikzlibrary{shapes.geometric}
\usetikzlibrary{calc}

\newcommand{\circled}[1]{\textcircled{\raisebox{-0.5pt}{\scriptsize #1}}}

\definecolor{Gray}{gray}{0.85}

\title{EFFICIENT DEEP ACOUSTIC ECHO SUPPRESSION \\ WITH CONDITION-AWARE TRAINING}

\name{Ernst Seidel$^{\ast}$, Pejman Mowlaee$^{\circ}$, Tim Fingscheidt$^{\ast}$}
\address{$^{\ast}$Institute for Communications Technology,
	Technische Universität Braunschweig\\
	Schleinitzstraße 22,
	38106 Braunschweig, Germany\\ $^{\circ}$GN Audio A/S,
	Lautrupbjerg 7,
	2750 Ballerup, Denmark}
\begin{document}

\ninept
\maketitle

\begin{sloppy}

\begin{abstract}
  The topic of deep acoustic echo control (DAEC) has seen many approaches with various model topologies in recent years. Convolutional recurrent networks (CRNs), consisting of a convolutional encoder and decoder encompassing a recurrent bottleneck, are repeatedly employed due to their ability to preserve nearend speech even in double-talk (DT) condition. However, past architectures are either computationally complex or trade off smaller model sizes with a decrease in performance.
  We propose an improved CRN topology which, compared to other realizations of this class of architectures, not only saves parameters and computational complexity, but also shows improved performance in DT, outperforming both baseline architectures FCRN and CRUSE. Striving for a condition-aware training, we also demonstrate the importance of a high proportion of double-talk and the missing value of nearend-only speech in DAEC training data. Finally, we show how to control the trade-off between aggressive echo suppression and near-end speech preservation by fine-tuning with condition-aware component loss functions.
\end{abstract}


\section{Introduction}
\label{sec:intro}

% \begin{itemize}
%     \item General Intro AEC + DAEC
%     \item Architectures -\> CRN
%     \item Efficient Networks
%     \item Condition-Aware
% \end{itemize}

The topic of acoustic echo control (AEC) has seen many scientific advancements in the recent years. While AEC is usually still conducted using classical approaches such as linear filters \cite{SpeexDSP,enzner_vary_fdaf}, more and more works have shown the tremendous advantage of employing deep neural networks, either as postfilters \cite{Valin_AEC,Zhang2022a}, hybrid systems \cite{Revach2021}, or as fully machine-learned echo cancellers \cite{westhausen21_icassp, seidel21_interspeech, Braun2022}.
% ,Peng2021

One of the more prominent architectural choices for deep AEC is the convolutional recurrent network (CRN) \cite{seidel21_interspeech, Wang2022, Braun2022}. These models consist of a convolutional encoder and decoder encompassing a recurrent bottleneck. Such models are capable at suppressing undesired components while preserving good speech quality \cite{Strake2020}. This makes them a great choice for AEC, as nearend (NE) speech preservation can be as important as aggressive echo suppression, wherever a postfilter is employed to remove residual echo alongside noise \cite{seidel21_interspeech, Braun2022}.
While CRNs, especially their fully convolutional variants \cite{seidel21_interspeech}, are great at preserving NE speech, they often exhibit high computational complexity and large network size. Recent proposals of more efficient CRNs \cite{Braun2022} report a less than optimal performance of the trained AEC as a stand-alone system.

% Another important aspect of AEC is the proportion of the three conditions double-talk (DT), farend single-talk (STFE), and nearend single-talk (STNE) in the training data. While a higher focus on DT as the most complex condition is common, the overall distribution varies and publications in the field of AEC rarely motivate their overall condition distribution.
The proportion of the three conditions double-talk (DT), farend single-talk (STFE), and nearend single-talk (STNE) in the AEC training data is also important. Many models are trained on data from the Microsoft AEC Challenge~\cite{cutler2022AEC} or on datasets of varying condition proportions \cite{Wang2022, Braun2022}, but their specific proportion choice is rarely motivated. Some authors propose losses sensitive to the microphone signal, e.g., using echo-aware loss weights per time-frequency bin \cite{Zhang2022a} or separate condition-specific losses \cite{pfeifenberger21_interspeech}.
% Some authors also propose condition-aware loss formulations, e.g., echo-aware loss weighting \cite{Zhang2022a}, to enhance their model's overall performance.

In this work, we adapt the {\tt FCRN15} noise reduction model~\cite{Strake2022} to acoustic echo suppression (AES) and propose modifications, {\it greatly decreasing computational complexity and parameter count}. We furthermore conduct ablation studies on the impact of the {\it minibatch condition splits (MCSs)} between the different conditions DT, STFE, and STNE. To improve the condition-aware training further, we propose {\it a single} condition-aware loss allowing to trade off between NE speech preservation and echo suppression {\it on a file basis}.

% The remainder of this paper will provide model and method descriptions in Section 2, experimental setup and evaluation in Section 3 and conclusions in Section 4.
The remainder of this paper is structured as follows: Section 2 introduces the processing framework, baseline architectures, proposed modifications and the condition-aware training concept. The datasets, training details and conducted experiments are presented in Section 3. Section 4 provides conclusions.

\section{System Overview and Proposed Method}
\label{sec:method}

\subsection{Processing Framework and Baseline Models}
\label{ssec:subhead}

All models described in this paper are implemented in the same processing framework to ensure a fair performance comparison. Our available input signals are the reference FE signal $x(n)$ and the microphone signal $y(n) = s(n) + n(n) + d(n)$ with NE speech $s(n)$, background noise $n(n)$, and echo $d(n) = f_\mathrm{NL}\left(x(n)\right) * h(n)$, with loudspeaker nonlinearity $f_\mathrm{NL}(\cdot)$ and room impulse response $h(n)$. All signals are sampled at $16$\,kHz. The signals $x(n)$ and $y(n)$ are split into frames of $N_T = 424$ samples and a frame shift of $50$\,\%, which are then subject to a square root Hann window and zero-padded so that a $K=512$-point DFT can be applied. The resulting frequency-domain representations $X_\ell(k)$ and $Y_\ell(k)$ with frame index $\ell$ and frequency bin index \mbox{$k \in \mathcal{K} = \{0,1,...,K/2\}$} are used as inputs to our DNN models. Both real and imaginary part form separate input channels and are zero-padded to $L\ge K/2+1$ to create a feature count divisible by the product of all encoder strides. The models are trained to estimate a mask $M_\ell(k)$ which is applied to $Y_\ell(k)$ after gain amplitude compression according to
\vspace{-1mm}
\begin{equation}
    E_\ell(k) = Y_\ell(k) \cdot \tanh(|M_\ell(k)|) \cdot\frac{M_\ell(k)}{|M_\ell(k)|}.
    \label{eq:mask}
    \vspace{-1mm}
\end{equation}
The enhanced signal $E_\ell(k)$ is transformed back into the time domain using a $K$-point IDFT, square-root Hann-windowed ($N_T$ samples window length), and subject to overlap-add (OLA), resulting in the time sequence $e(n)$.

We compare our proposed model to two baselines architectures: The {\tt FCRN} \cite{seidel21_interspeech} as a large, high-quality model, and the {\tt CRUSE DAEC-64} architecture \cite{Braun2022} including input amplitude compression (further simply {\tt CRUSE}), as an efficient model with low parameter count and computational complexity.

\subsection{Proposed Efficient Architecture}
\label{ssec:architecture}

\begin{figure}[t]
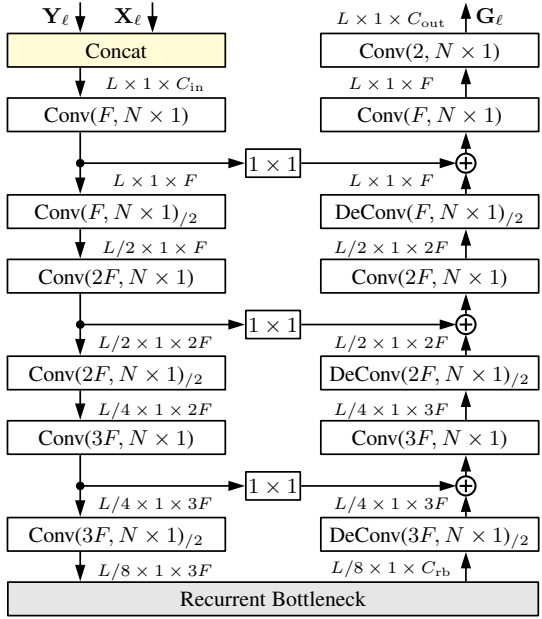

	% \centering
	
    \includestandalone[width=1.0\columnwidth, mode=image|tex]{fig/xCRN_long}
	\vspace{-6mm}
	\caption{The general \texttt{CRN} topology employed for \texttt{FCRN15}, \texttt{gGCRN16}, and all in-between models, depicted using their respective recurrent bottleneck (with output channel dimension C$_\mathrm{rb}$).}
	\label{fig:xCRN}
	\vspace{-5mm}
\end{figure}

% Our proposed model, created in the {\tt Tensorflow2} framework \cite{}, is based on the FCRN model \cite{}. 
While the {\tt FCRN} yields good quality, it has a high parameter count and computational complexity. As it is used in various fields of speech enhancement, more efficient variants have been developed, but not yet evaluated on the task of AES. We adapt the {\tt FCRN15} structure described in \cite{Strake2022, Strake2020a}, shown in Fig.\,\ref{fig:xCRN}. Every other encoder layer features a stride of 2 (labelled as $_{/2}$) to reduce the feature map size, which is restored using deconvolutional layers in the decoder. Depth-wise convolutions are employed in the skip connection paths. The recurrent bottleneck (grey box) is represented by two sequential ConvLSTM layers of kernel size $(N \times 1)$ and kernel count $F$, resulting in $C_\mathrm{rb}=F$.
For the use in AES, we concatenate real and imaginary parts of both input signals $X_\ell(k)$ and $Y_\ell(k)$ before passing them to the first layer ($C_\mathrm{in}=4$). We choose $L=264$, $N=12$, and $F=32$. This model already decreases the required floating point operations per second (FLOPS) by a factor of six compared to the {\tt FCRN}, but is still quite slow in training and inference. We propose further improvements to the {\tt FCRN15}:

\circled{1} \textbf{Grouped GRUs}: Replacing the \mbox{ConvLSTM} layers in the bottleneck of the {\tt FCRN15} with GRU layers massively improves training speed, but considerably increases the parameter count due to the fully connected nature of GRUs. To counteract this, we add a convolutional layer of kernel count $F$ in the beginning of the \mbox{bottleneck} and split its output feature maps between multiple parallel GRUs. This first model uses two consecutive layers of $8$ and $6$ grouped GRUs (gGRUs), respectively, and representation rearrangement~\cite{Tan2020}. A convolutional layer with kernel count $3F$ after the gGRU layers restores the input dimensions.

\circled{2} \textbf{Kernel count}: We found that increasing the kernel count from $F = 32$ to $F = 40$ improved the model's performance notably at the cost of a moderately increased computational complexity. To split our channels in the recurrent bottleneck uniformly, the first gGRU layer now consists of $10$ (instead of $8$) GRUs.

\circled{3} \textbf{Kernel size}: We found that a reduction of the kernel size did not negatively affect the models performance down to $N=3$, which reduces the computational complexity of our encoder and decoder by $75$\,\%.

\circled{4} \textbf{Input compression:} We adopt the input compression from \cite{Braun2022} defined as 
$Y^\mathrm{compressed}_\ell(k) = |Y_\ell(k)|^c \cdot e^{j\varphi_{Y,\ell}(k)}$
with $c=0.3$ for both model inputs $Y_\ell(k)$ and $X_\ell(k)$. Note that this is only applied to the encoder inputs, and that we do not compress or decompress any signals of the masking operation (\ref{eq:mask}). We found this practice to improve our model's performance in double-talk.

\circled{5} \textbf{GRU layer simplification}: As a last step to create our proposed model, further labelled as {\tt gGCRN16}, we remove the second gGRU layer, causing only slightly reduced performance. The resulting recurrent bottleneck is depicted in Fig.\,\ref{fig:gGCRN16}. This step greatly reduces our model's parameter count. Compared to the {\tt FCRN15}, we reduce FLOPS by 70\%. Compared to {\tt CRUSE}, we require 30\% less parameters and 15\% less FLOPS.

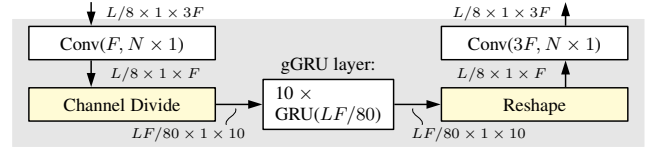
\begin{figure}[t]
	\centering
	\vspace{-1mm}
    	\usetikzlibrary{dsp}
	\usetikzlibrary{chains}
	\usetikzlibrary{shapes.geometric}
	\usetikzlibrary{calc}

\begin{tikzpicture}[scale = .7]

\tikzstyle{arrow} = [thick,->,>=stealth]
\pgfmathsetmacro{\rootx}{0}
\pgfmathsetmacro{\rooty}{0}

\pgfmathsetmacro{\topconvs}{-1}
\pgfmathsetmacro{\convsep}{1.5}

\pgfmathsetmacro{\leftrow}{-3.0}
\pgfmathsetmacro{\rightrow}{8.5}
\pgfmathsetmacro{\middlerow}{(\rightrow+\leftrow)/2}
\pgfmathsetmacro{\boxheight}{1.8}
\pgfmathsetmacro{\boxwidth}{10}
\pgfmathsetmacro{\quadboxwidth}{34}

	% \fill [white] (-3,-0.1) rectangle (10,-4.6);
	
	\draw   (\leftrow,\rooty+\convsep-1.5)   coordinate    (NoisyInput)    {}
	       (\rightrow,\rooty+\convsep-1.5)  coordinate    (MaskOut)    {}
              (\middlerow,{\topconvs-0.65*\convsep}) node    (backdrop)     [dspfilter, minimum width=\quadboxwidth em, minimum height=7.0 em, fill=black!10, draw=white] {}
	       (\leftrow+.75,{\topconvs-0*\convsep}) node    (bottIn)     [dspfilter, minimum width=\boxwidth em, minimum height=\boxheight em, fill=white] {Conv($F, N\times 1$)}
	       (\rightrow-0.75,{\topconvs-0*\convsep}) node    (bottOut)     [dspfilter, minimum width=\boxwidth em, minimum height=\boxheight em, fill=white] {Conv($3F, N\times 1$)}
		   (\leftrow,{\topconvs-(0.55*\convsep)})   coordinate    (bottIn_)    {}
		   (\rightrow,{\topconvs-(0.55*\convsep)})   coordinate    (bottOut_)    {}
              (\leftrow+.75,{\topconvs-1*\convsep}) node    (clstm1)     [dspfilter, minimum width=\boxwidth em, minimum height=\boxheight em, fill=yellow!20] {Channel Divide}
	       (\rightrow-0.75,{\topconvs-1*\convsep}) node    (clstm2)     [dspfilter, minimum width=\boxwidth em, minimum height=\boxheight em, fill=yellow!20] {Reshape}
              ({\middlerow},{\topconvs-1*\convsep}) node    (rr)     [dspfilter, minimum width=0.7*\boxwidth em, minimum height=1.5*\boxheight em, fill=white] {}
              ({\middlerow},{\topconvs-1.55*\convsep}) node (text) 
              % [label=10 $\times$ \\ gGRU($LF/80$)] 
              [label=\vtop{\hbox{\strut{10 $\times$}} \hbox{GRU($LF/80$)}}]
              {};
     %          ({\middlerow},{\topconvs-2*\convsep}) node    (rr)     [dspfilter, minimum width=\boxwidth em, minimum height=\boxheight em, fill=white] {10 $\times$ gGRU($LF/80$)}
     %          (\leftrow,{\topconvs-(2*\convsep)})   coordinate    (clstm1_)    {}
		   % (\rightrow,{\topconvs-(2*\convsep)})   coordinate    (clstm2_)    {};

% \vtop{\hbox{\strut{\tt FCRN}} \hbox{after \cite{}}}

	%%% labels for named Signals
	% \draw  ([xshift=-0mm, yshift=-3mm] NoisyInput)   node    (inputTop)       [label=above:$\mathbf{Y}_{\ell}(k)$] {};
	% \draw  (MaskOut)   node    (inputTop)       [label=right:$\mathbf{G}_{\ell}(k)$] {};
	
    \draw ([xshift=8mm, yshift=-2mm] bottIn.north) coordinate (bottIn_data_in);
    \draw ([xshift=8mm, yshift=-2mm] clstm1.north) coordinate (clstm1_data_in);
    \draw ([xshift=8mm, yshift=2mm] clstm1.south) coordinate (clstm1_data_out);
    \draw ([xshift=-8mm, yshift=-2mm] clstm2.north) coordinate (clstm2_data_out);
    \draw ([xshift=-8mm, yshift=2mm] clstm2.south) coordinate (clstm2_data_in);
	\draw ([xshift=-8mm, yshift=-2mm] bottOut.north) coordinate (bottOut_data_out);	

    \draw  ([yshift=-2mm] rr.north)   node    (out_data_node)       [label=above:gGRU layer:] {};

    \draw  (bottIn_data_in)   node    (out_data_node)       [label=above:\scriptsize $L/8 \times 1 \times 3F$] {};
    \draw  (clstm1_data_in)   node    (out_data_node)       [label=above:\scriptsize $L/8 \times 1 \times F$] {};
    \draw  ([xshift=+8mm, yshift=+0.0mm]clstm1_data_out)   node    (clstm1_out)       [label=below:\scriptsize $LF/80 \times 1 \times 10$] {};
    \draw  ([xshift=-8mm, yshift=+0.0mm]clstm2_data_in)   node    (clstm1_out)       [label=below:\scriptsize $LF/80 \times 1 \times 10$] {};
    \draw  (clstm2_data_out)   node    (out_data_node)       [label=above:\scriptsize $L/8 \times 1 \times F$] {};
	\draw  (bottOut_data_out)   node    (out_data_node)       [label=above:\scriptsize $L/8 \times 1 \times 3F$] {};
	
	% \draw ([xshift=-35mm] quadconv1.north) coordinate (quadconv1.in1);
	% \draw ([xshift=35mm] quadconv1.north) coordinate (quadconv1.out2);
	% \draw ([xshift=-35mm] quadconv1.south) coordinate (quadconv1.out1);
	% \draw ([xshift=35mm] quadconv1.south) coordinate (quadconv1.in2);
	
	% \draw ([xshift=-35mm] quadconv2.north) coordinate (quadconv2.in1);
	% \draw ([xshift=35mm] quadconv2.north) coordinate (quadconv2.out2);
	% \draw ([xshift=-35mm] quadconv2.south) coordinate (quadconv2.out1);
	% \draw ([xshift=35mm] quadconv2.south) coordinate (quadconv2.in2);
	
	% \draw ([xshift=-35mm] quadconv3.north) coordinate (quadconv3.in1);
	% \draw ([xshift=35mm] quadconv3.north) coordinate (quadconv3.out2);
	% \draw ([xshift=-35mm] quadconv3.south) coordinate (quadconv3.out1);
	% \draw ([xshift=35mm] quadconv3.south) coordinate (quadconv3.in2);

    \draw ([xshift=-7.5 mm] bottIn.north) coordinate (bottIn.in);
	\draw ([xshift=-7.5mm] bottIn.south) coordinate (bottIn.out);
	
	\draw ([xshift=+7.5mm] bottOut.north) coordinate (bottOut.out);
	\draw ([xshift=+7.5 mm] bottOut.south) coordinate (bottOut.in);
	
	\draw ([xshift=-7.5 mm] clstm1.north) coordinate (clstm1.in);
	\draw ([xshift=-7.5mm] clstm1.south) coordinate (clstm1.out);
	
	\draw ([xshift=+7.5mm] clstm2.north) coordinate (clstm2.out);
	\draw ([xshift=+7.5 mm] clstm2.south) coordinate (clstm2.in);
	
	\draw ([xshift=+7.5mm] out_conv.north) coordinate (out_conv.out);
	\draw ([xshift=+7.5 mm] out_conv.south) coordinate (out_conv.in);

	\begin{scope}[start chain]
	
	\chainin (NoisyInput);

    \chainin (bottIn.in)        [join=by dspconn];
    \chainin (bottIn.out) ;
	\chainin (clstm1.in) 		[join=by dspconn];
 
	\chainin (clstm1);
	% \chainin (clstm1_) 		[join=by dspline];
    \chainin (rr) 		    [join=by dspconn];
	% \chainin (clstm2_) 		[join=by dspline];
	\chainin (clstm2) 		[join=by dspconn];
	
	\chainin (clstm2.out);
    \chainin (bottOut.in)        [join=by dspconn];
    \chainin (bottOut.out) ;

	\chainin (MaskOut) 	[join=by dspconn];

    \end{scope}

    \draw ([xshift=-6.5mm, yshift=-1.0mm] rr.west) coordinate (wiggle_A1);
	\draw ([xshift=+2.5mm, yshift=-0.5cm] clstm1.east) coordinate (wiggle_A2);
    \draw (wiggle_A1) to [out=-70, in=+110] (wiggle_A2);

    \draw ([xshift=+4.5mm, yshift=-1.0mm] rr.east) coordinate (wiggle_B1);
	\draw ([xshift=-4.5mm, yshift=-0.5cm] clstm2.west) coordinate (wiggle_B2);
    \draw (wiggle_B1) to [out=-110, in=+70] (wiggle_B2);
	
\end{tikzpicture}
    \vspace{-6mm}
	\caption{The \texttt{gGCRN16} recurrent bottleneck ($C_\mathrm{rb}=3F$) using a grouped GRU strategy with 10 parallel layers ($F=40$).}
	\label{fig:gGCRN16}
	\vspace{-4mm}
\end{figure}

\subsection{Proposed Condition-Aware Training and Losses}
\label{ssec:condition}

To implement a condition-aware training, we use specific minibatch condition splits (MCSs). This means that each minibatch contains a given number of sequences for each of the conditions double-talk (DT), farend single-talk (STFE), and nearend single-talk (STNE). By changing the proportion of each condition in a batch, we can better control our model's training behaviour. The representation of all used training conditions in each batch mitigates the danger of detrimental weight updates caused by an imbalanced MCS, e.g., a minibatch only containing STFE and no echo.

The knowledge of each minibatch entry's specific condition can also be leveraged to apply a condition-aware loss formulation which makes use of the white-box speech component \mbox{$\tilde{S}_\ell(k) = G_\ell(k) \cdot S_\ell(k)$} and residual echo component \mbox{$\tilde{D}_\ell(k) = G_\ell(k) \cdot D_\ell(k)$} by applying our estimated mask $G_\ell(k)$ to the respective microphone signal components.
% \begin{align}
% \tilde{S}_\ell(k) & = G_\ell(k) \cdot S_\ell(k) \\
% \tilde{D}_\ell(k) & = G_\ell(k) \cdot D_\ell(k).
% \label{eq:wb_comp}
% \end{align}
Using these components in the time domain, the condition-aware loss for each minibatch entry $b$ is computed as
\begin{align}
\begin{split}
J^\mathrm{CA}_b = (1-\alpha_b-\beta_b) J^{\mathrm{logMSE}}_b \big({e}(n), {s}_b(n) + {n}_b(n)\big) \\  
\ \ + \alpha_b J^{\mathrm{logMSE}}_b \big(\tilde{s}_b(n), {s}_b(n)\big) + \beta_b J^{\mathrm{logMSE}}_b \big(\tilde{d}_b(n), {0}\big),
\end{split}
\label{eq:cda_loss}
\end{align}
where $\alpha_b$ and $\beta_b$ are weighting factors of their respective loss terms. These weights are specific to each individual batch entry based on its condition (DT, STFE, STNE). The loss
% \vspace{-1mm}
\begin{align}
    {J}^{\mathrm{logMSE}}_b(\hat{{z}}_b(n),\,\, \overline{z}_b(n)) & = 10\!\cdot\!\log \left(\sum_{n \in \mathcal{N}} \big(\hat{{z}}_b(n),\,\, \overline{z}_b(n)\big)^2  \right) \label{eq:tlmse}
    % \vspace{-1mm}
\end{align}
computes the log MSE over the entire time sequence $(n \in \mathcal{N})$ with target $\overline{z}_b(n)$ and its estimate $\hat{z}_b(n)$. Note that the target for the last term in (\ref{eq:cda_loss}) is a sequence of zeros, as we want our residual echo power to be minimized. Such a loss formulation can be used either directly in the training or alternatively in a fine-tuning step. We focus on the application in fine-tuning, since both methods lead to similar results, but fine-tuning allows for faster ablations and proved more stable for higher values of $\alpha_b$ and $\beta_b$.
 
\section{Experimental Evaluation and Discussion}
\label{sec:setup}

\subsection{Dataset and Metrics}
\label{ssec:dataset}

The \textit{training dataset} $\mathcal{D}_{\mathrm{train}}$ uses speakers from the CSTR-VCTK corpus~\cite{VCTK} to generate NE and FE utterances. Each speaker can appear as NE or FE to discourage the trained model from overfitting to speakers. To generate the echo signal, first a scaled error function (SEF)~\cite{Zhang2020c,Klippel2006} defined as 
$f_\mathrm{SEF}(x(n)) = \int_{0}^{x} exp\left({\frac{x^2(n)}{2\mu^2}}\right) dx$
is applied to the FE reference signal $x(n)$, with $\mu$ randomly chosen from $\{0.5, 1, 10, 999\}$. This simulates possible non-linear distortions caused by the loudspeaker. The resulting distorted signal is convolved with an impulse response (IR) generated via the image method \cite{image_method}. 
The signal-to-echo ratio (SER) of each audio file is randomly sampled from a continuous range of $[-12.4,22.4]$\,dB. For background noise we use random cuts from the DEMAND~\cite{Thiemann2013} and QUT-NOISE~\cite{QUT} databases at a signal-to-noise ratio (SNR) in the range of $[-2.4, 32.4]$\,dB, with a $10$\% chance of a file to be noiseless. We generate $9500$ files of $10$\,s length, of which $1000$ files are only used for validation and learning rate control during training. Between training epochs, the speech, noise, and echo components of $\mathcal{D}_{\mathrm{train}}$ (except the validation split) are reshuffled with new SER and SNR values.

For preliminary evaluations we use a \textit{development dataset} $\mathcal{D}_{\mathrm{dev}}$ of 200 files. Speakers are drawn from a speaker subset of the CSTR-VCTK corpus disjoint to the training speakers. Nonlinear distortions are again simulated via $f_\mathrm{SEF}(x(n))$, but with distinct $\alpha$ values sampled from $\{0.2, 0.4, 1.5, 12, 999\}$. The image method is used for IR generation with the same parameter setup as in $\mathcal{D}_{\mathrm{train}}$, but different random seeding. SER values are sampled from discrete values in $\{-10, -5, ..., 20\}$\,dB. Noise is cut from the ETSI database~\cite{ETSI2008}. The SNR is chosen from $\{0, 5, ..., 30\}$\,dB.
Contrary to $\mathcal{D}_{\mathrm{train}}$, each file contains distinct sections of all three conditions in the order \mbox{STFE, STNE, DT}. All sections are of $8$ to $12$\,s length.

The \textit{test dataset} $\mathcal{D}_{\mathrm{test}}$ is created from speakers of the TIMIT speech corpus~\cite{TIMIT}. The arctan nonlinearity function~\cite{Jung2013, Zoellzer2003} defined as
$f_\mathrm{arctan}(x(n)) = {\arctan{(\alpha \cdot x(n))}}/{\alpha}$ models the loudspeaker nonlinearites, with \mbox{$\alpha=10^{-4}$}. For IR generation we use the image method, but adjust its parameters to include bigger room dimensions, more variance in speaker / loudspeaker distance and position, and different reflection coefficients. Noise is cut from the remaining ETSI files of environments unseen in the training. The SER is chosen from $\{-9, -6, ..., 9\}$\,dB, the SNR from $\{5, 8, ..., 20\}$\,dB. We generate 200 files in the same fashion as for $\mathcal{D}_{\mathrm{dev}}$.

All three conditions are evaluated on their own subset of metrics. Metrics used in this work include PESQ~\cite{ITU_P862.2_Corr}, the ERLE metric 
% \cite{Vary2006}
% defined as $ERLE(n)=10\mathrm{log}_{10}(d^2(n)/(d^2(n)-\hat{d}(n))^2)$,
% for which a first-order IIR smoothing filter with factor $\alpha=0.99$ is applied to the  signals $d^2(n)$ and $d_\Delta^2(n)=(e(n)-n(n))^2$,
after \cite{Vary2006} with a first-order IIR smoothing filter ($\alpha=0.99$) and the STFE echo estimate defined as $\hat{d}(n)=e(n)-n(n)$,
and the \mbox{AECMOS} metrics~\cite{purin2021aecmos}.
To observe the trade-off between NE speech preservation and echo suppression in the fine-tuning ablations, we use the black-box component metrics PESQ$_\mathrm{BB}$ and ERLE$_\mathrm{BB}$ as described in \cite{fingscheidt_signalseparation, seidel21_interspeech}.

\begin{table}[t]
	\setlength{\tabcolsep}{.35em}
    \vspace{-2mm}
	\caption{Effects of \textbf{condition-aware} \textbf{training} on {specific minibatch condition splits (MCSs)} DT/STFE/STNE, evaluated on the \textbf{dev set} $\mathcal{D}_{\mathrm{dev}}$. Best results are {bold}, second best {underlined}.}
	\vspace{0.5mm}
    \centering
	\newcolumntype{R}{>{\raggedleft\arraybackslash}X}
\newcolumntype{C}{>{\center\arraybackslash}X}
\newcommand{\bftab}{\fontseries{b}\selectfont}

\begin{tabular}{cc ccc c c} 

	\toprule
{\multirow{2}{*}{\rotatebox[origin=c]{90}{\textbf{Model}}}} & \multirow{2}{*}{\textbf{MCS}} & \multicolumn{3}{c}{\textbf{Double-Talk}} & \multicolumn{1}{c}{\textbf{STFE}} & \multicolumn{1}{c}{\textbf{STNE}}\\

\cmidrule(lr){3-5}
\cmidrule(lr){6-6}
\cmidrule(lr){7-7}

& & PESQ & DT\,Other & DT\,Echo & ST\,Echo & ST\,Other\\
\cmidrule(lr){1-7}

 \multicolumn{2}{c}{Unprocessed} & 1.84 & 4.03 & 2.14 & 1.92 & 3.33 \\
 \cmidrule(lr){1-7}
 {\multirow{6}{*}{\rotatebox[origin=c]{90}{{\tt gGCRN16}}}} 
 & (rand)  &  1.99 & 2.80 & 4.19 & \underline{4.48} & 3.31 \\
 & (8/8/0)  & 2.33  & 3.13  & \underline{4.42}  & \textbf{4.50}  & {3.33} \\
 & (12/4/0)  & 2.44 & 3.25 & 4.41 & 4.46 & {3.32}\\
 & (13/2/1)  & {2.53} & 3.35 & \underline{4.42} & {4.47} & {3.32}\\
 & (15/1/0)  & \underline{2.61} & {3.41} & \textbf{4.45} & 4.46 & {3.33}\\
 & (16/0/0) & \underline{2.61} & \textbf{3.44} & \underline{4.42} & 4.44 & {3.33}\\

 \cmidrule(lr){1-7}
 % {\multirow{3}{*}{\rotatebox[origin=c]{90}{{\tt FCRN}}}}
 {\multirow{3}{*}{\rotatebox[origin=c]{90}{
\vtop{\hbox{\strut{\tt FCRN}} \hbox{after \cite{seidel21_interspeech}}}
 }}} 
 & (rand) &  2.42 & 3.26 & 4.35 & 4.51 & 3.32\\
 & (8/8/0) & 2.38 & 3.19 & 4.40 & \textbf{4.50} & 3.33\\
 & (16/0/0) &\textbf{2.63} & \underline{3.43} & 4.40 & 4.46 & \textbf{3.35} \\
 \cmidrule(lr){1-7}
 {\multirow{3}{*}{\rotatebox[origin=c]{90}{
\vtop{\hbox{\strut{\tt CRUSE}} \hbox{after \cite{Braun2022}}}
 }}} 
 & (rand) & 2.26 & 3.21 & 4.29 & 4.45 & 3.32\\
 & (8/8/0) & 2.25 & 3.18 & 4.39 & \textbf{4.50} & \textbf{3.35} \\
 & (16/0/0) & 2.52 & 3.38 & 4.30 & 4.42 & \underline{3.34}\\

\bottomrule
\end{tabular}
    \vspace{-3mm}
	\label{tab:results_dist_dev}
\end{table}

\begin{table*}[t]
	\setlength{\tabcolsep}{.35em}
	\caption{Proposed modifications to the {\tt FCRN} architecture up to the {\tt gGCRN16} as described in Section \ref{ssec:architecture}, evaluated on the \textbf{dev set} $\mathcal{D}_{\mathrm{dev}}$. All models are trained on \mbox{MCS (16/0/0)}. Best results are {bold}, second best {underlined}. \colorbox{Gray}{Our proposed model architecture is highlighted in grey.}}
	\vspace{0.5mm}
    \centering
	\newcolumntype{R}{>{\raggedleft\arraybackslash}X}
\newcolumntype{C}{>{\center\arraybackslash}X}
\newcommand{\bftab}{\fontseries{b}\selectfont}

% \definecolor{Gray}{gray}{0.85}

% \newcolumntype{a}{>{\columncolor{Gray}}c}

% \newcommand{\circled}[1]{\textcircled{\raisebox{-0.5pt}{\scriptsize #1}}}

% {\vtop{\hbox{\strut{\textbf{\#FLOPS/}}} \hbox{\strut{\textbf{ frame\phantom{a.}}}}}}

\begin{tabularx}{\linewidth}{l R R c ccc cc cc} 

	\toprule
\multirow{2}{*}{\textbf{Model}} & \multirow{2}{*}{\textbf{\# Parameters}} & \multirow{2}{*}{\textbf{\#FLOPS}} & & \multicolumn{3}{c}{\textbf{Double-Talk}} & \multicolumn{2}{c}{\textbf{STFE}} & \multicolumn{2}{c}{\textbf{STNE}}\\

% \multirow{2}{*}{\vtop{\hbox{\strut{\textbf{\#FLOPS/}}} \hbox{\strut{\textbf{ frame\phantom{a.}}}}}}
% \multirow{2}{*}{\textbf{\#FLOPS/ frame\phantom{a.}}}

\cmidrule(lr){5-7}
\cmidrule(lr){8-9}
\cmidrule(lr){10-11}

& & & & PESQ & DT Other & DT Echo & ERLE & ST Echo & PESQ & ST Other\\
\cmidrule(lr){1-11}

 Unprocessed & - & - & & 1.84 & 4.03 & 2.14 & - & 1.92 & 3.11 & 3.33 \\

 \cmidrule(lr){1-11}
 
 % {\tt FCRN} after \cite{}  & (rand) & 3.7 M & 881 \phantom{.} M &  \\
 % {\tt FCRN}  & (8/8/0) & & &  \\
 {\tt FCRN} after \cite{seidel21_interspeech} & 3.7 M & 12,840 M & & \underline{2.63} & \underline{3.43} & 4.40 & \underline{16.26} & 4.46 & \underline{3.14} & \underline{3.35}\\
  
% \cmidrule(lr){1-11}

{\tt FCRN15}  after \cite{Strake2022} & \bftab{1.0 M} & 2,011 M & & 2.59 & 3.30 & 4.35 & 16.20 & \textbf{4.50} & 3.13 & 3.34\\
 + \circled{1} (gGRU for ConvLSTM) & {2.3 M} & 772 M & & 2.61 & 3.32 & 4.41 & \underline{16.26} & \underline{4.48} & 3.13 & 3.34\\
 + \circled{2} (increased $N=40$) & {3.4 M} & 1,180 M & & 2.62 & 3.36 & 4.41 & \underline{16.26} & \underline{4.48} & 3.13 & 3.33\\
 + \circled{3} (reduced $k=3$) & {3.1 M} & \underline{641 M} & & 2.61 & 3.37 & \underline{4.44} & 16.20 & 4.46 & \textbf{3.15} & \textbf{3.36}\\
 + \circled{4} (input amplitude compression) & {3.1 M} & \underline{641 M} & & \textbf{2.66} & \textbf{3.44} & \textbf{4.47} & \textbf{16.63} & 4.47 & 3.12 & 3.34\\
 \rowcolor{Gray}
 + \circled{5} (= {\tt gGCRN16}) \quad (proposed) & \underline{1.3 M} & \bftab{583} M & & {2.61} & \textbf{3.44} & {4.42} & 16.23 & 4.44 & 3.12 & {3.33}\\
 {\tt CRUSE} after~\cite{Braun2022} & 1.9 M & {685 M} & & 2.52 & 3.38 & 4.30 & 15.64 & 4.42 & 3.13 & 3.34\\
 % \cmidrule(lr){1-11}
  % {\tt CRUSE-AEC} & (8/8/0) & & & 1.97& 3.73 & 4.29 & 12.21 & 4.24 & 2.75 & 3.51\\
  % {\tt CRUSE-AEC} & (rand) & & & 1.80 & 3.29 & 4.29 & 12.16 & \textbf{4.51} & 2.33 & 2.99\\
 
\bottomrule
\end{tabularx}
    \vspace{-3mm}
	\label{tab:results_arch_dev}
\end{table*}

\subsection{Training Details}
\label{ssec:training}

% Publications on the field of learned AEC use various often not clearly motivated compositions of training conditions and usually combine the training task with joint noise reduction (NR). There are only few examples where an AEC-only model is explicitly trained \cite{,}. We choose a random MCS concerning DT / STFE / STNE as starting point for our experiments. Although we are aware that such an equal distribution of the three distributions is not standard in literature, it serves as a good reference for the performance difference achieved through condition-aware training.

We use the Adam optimizer~\cite{adam_optimizer} in its standard configuration for model training. Apart from the fine-tuning ablations on condition-aware losses, all models are trained on loss (\ref{eq:cda_loss}) with $\alpha_b = \beta_b = 0$. The batch size is set to $16$ with a backpropagation-through-time (BPTT) unrolling sequence length of $200$ frames. 
%Between epochs, the components of $\mathcal{D}_{\mathrm{train}}$ are reshuffled with new SER and SNR values.

The initial learning rate (LR) is set to $10^{-4}$, which is halved after 4 epochs without loss improvement on the validation split of $\mathcal{D}_{\mathrm{train}}$. The training is stopped after 100 epochs, if the validation loss does not improve for 10 consecutive epochs, or if the LR drops below $10^{-5}$. The fine-tuning ablations are trained for 30 epochs using the same LR strategy with an initial LR of $2.5\cdot10^{-5}$ and minimal LR of $2.5\cdot10^{-6}$.
All models are trained in {\tt TensorFlow2}~\cite{Abadi2016}.

\subsection{Development Set Ablations}
\label{ssec:dev}

Table \ref{tab:results_dist_dev} shows the effects of different MCSs as the specific number of files for each processed minibatch (DT / STFE / STNE), evaluated on the {\tt gGCRN16} and for selected MCSs also on both baseline models {\tt FCRN}~\cite{seidel21_interspeech} and {\tt CRUSE}~\cite{Braun2022}. We slowly shift the MCS towards DT-only (16/0/0), as DT is the most challenging condition and therefore the most valuable in training. One additional experiment is conducted with MCS (8/8/0). This resembles the distribution of the synthetic dataset provided for the INTERSPEECH 2021 AEC Challenge, which used files consisting of $10$\,s of FE audio and zero-padded NE audio of $3$-$7$\,s length \cite{cutler2022AEC}.

{\it We can see that the DT condition metrics on $\mathcal{D}_{\mathrm{dev}}$ mostly improve with higher proportion of DT in the training set, while the STFE condition performance only slightly benefits from more STFE representation in training} (e.g., ST Echo from $4.44$ to $4.50$ MOS points on the {\tt gGCRN16}). While no model shows particularly good performance on the random MCS training, the {\tt gGCRN16} with its low computational complexity seems to be more sensible to this training configuration. Note that these findings also hold true on $\mathcal{D}_{\mathrm{test}}$. {\it Interestingly, the lack of STNE training data in (n/m/0) MCSs did not lead to any considerable degradation of the near-end speech during single-talk on neither $\mathcal{D}_{\mathrm{dev}}$ nor $\mathcal{D}_{\mathrm{test}}$.} 

Table \ref{tab:results_arch_dev} shows the proposed modifications to the {\tt FCRN} model as detailed in Section \ref{ssec:architecture}. We also include our implementation of the {\tt CRUSE} as a reference for an efficient architecture. Based on the findings of Table \ref{tab:results_dist_dev}, we train all models on MCS (16/0/0).

These results show the effectiveness of our architectural improvements. The performance stays largely consistent to the original architecture. Our proposed steps repair the initial performance drop of the {\tt FCRN15} architecture in DT while further reducing computational complexity (70\% less). We observe a slight performance degradation caused by the removal of the second gGRU layer in our final proposed {\tt gGCRN16} model, which we see as a reasonable trade-off for the huge savings in parameter count. Furthermore, we outperform the {\tt CRUSE} architecture in model size (-30\%), computational complexity (-15\%), and performance. Again, STNE performance remains at a high level with negligible variations.

\subsection{Test Set Results and Discussion}
\label{ssec:test}

\begin{figure}
    \vspace{-1.5mm}
    % \centering
    \hspace{-1.7em}%
    % \begin{subfigure}
        	\usetikzlibrary{dsp}
	\usetikzlibrary{chains}
	\usetikzlibrary{shapes.geometric}
	\usetikzlibrary{calc}

\pgfkeys{
    /prepare label/.style={
        /print label/\detokenize{#1}/.code={\ttfamily\detokenize{#1}}
    },
    /prepare label/.list={FCRN,FCRN15, gGCRN16, CRUSE, FDKF}
}

\begin{tikzpicture}

\begin{axis}[
width=.60*\columnwidth,
height=.60*\columnwidth,
legend cell align={left},
legend style={fill opacity=1.0, draw opacity=1, text opacity=1, draw=white!80!black},
log basis x={10},
tick align=outside,
tick pos=left,
xlabel={\#FLOPS},
% x labels style = {color=black}
xmajorgrids,
xmin=300000000, xmax=20000000000,
xmode=log,
tick style={color=black},
y label style={anchor=north, xshift=17mm, yshift=-3mm},
ylabel={\rotatebox[origin=c]{270}{{MOS}}},
ymajorgrids,
yminorgrids,
minor y tick num=4,
ymin=3.3, ymax=4.7,
extra y ticks={4.6},
extra y tick labels={
    {}
},
extra y tick style={%
    major tick length=3pt,
},
% ytick style={color=black},
extra x ticks={350000000},
extra x tick style={%
    every tick/.append style={white, line width=2.7pt},
    yshift=1pt,
    % every tick/.append style={red},
    grid=none,
},
extra x tick labels={
    {}
    % \rotatebox[origin=c]{270}{\rule{3mm}{$\approx$}}
},
]
\addplot[
        mark=star,
        line width=0.75pt,
        only marks,
        point meta=explicit symbolic,
        violet,
        nodes near coords={
            \pgfkeys{/print label/\pgfplotspointmeta/.try}
        },
        every node near coord/.append style={font=\small, anchor=south,yshift=-13pt}
    ]
        table[header=false,meta index=0,x index=1,y index=2]{
			FCRN_ 12840000000 3.94
        };
\addplot[
        mark=star,
        line width=0.75pt,
        only marks,
        point meta=explicit symbolic,
        blue,
        nodes near coords={
            \pgfkeys{/print label/\pgfplotspointmeta/.try}
        },
        every node near coord/.append style={font=\small, anchor=south,yshift=-13pt, inner sep=2}
    ]
        table[header=false,meta index=0,x index=1,y index=2]{
			FCRN15_ 2011000000 3.80
        };
\addplot[
        mark=star,
        line width=0.75pt,
        only marks,
        point meta=explicit symbolic,
        cyan,
        nodes near coords={
            \pgfkeys{/print label/\pgfplotspointmeta/.try}
        },
        every node near coord/.append style={font=\small, anchor=south,yshift=-18pt, inner sep=2}
    ]
        table[header=false,meta index=0,x index=1,y index=2]{
			gGCRN16_ 583000000 3.98	
        };
\addplot[
        mark=star,
        line width=0.75pt,
        only marks,
        point meta=explicit symbolic,
        magenta,
        nodes near coords={
            \pgfkeys{/print label/\pgfplotspointmeta/.try}
        },
        every node near coord/.append style={font=\small, anchor=south,yshift=-13pt, inner sep=2}
    ]
        table[header=false,meta index=0,x index=1,y index=2]{
			CRUSE_ 685000000 3.92
        };
\addplot[
        mark=star,
        line width=0.75pt,
        only marks,
        point meta=explicit symbolic,
        orange,
        nodes near coords={
            \pgfkeys{/print label/\pgfplotspointmeta/.try}
        },
        every node near coord/.append style={font=\small, anchor=south,yshift=3.5pt, xshift=17pt, inner sep=2}
    ]
        table[header=false,meta index=0,x index=1,y index=2]{
			FDKF 330000000 3.63
        };
\end{axis}

\begin{axis}[
width=.60*\columnwidth,
height=.60*\columnwidth,
  axis y line*=right,
  axis x line=none,
xmin=300000000, xmax=20000000000,
xmode=log,
  ymin=3.3, ymax=4.7,
  % y axis line style=black,
% y label style={anchor=north, xshift=23mm, yshift=-81mm},
% ylabel={\rotatebox[origin=c]{270}{{MOS}}},
  %ylabel={DeltaSNR},
  every tick label/.append style={ticks=none, font=\big},
  legend pos=south east  
]
\addlegendimage{only marks, mark=o, black} \addlegendentry{DT\,Echo}
\addlegendimage{only marks, mark=star, black}\addlegendentry{DT\,Other}
\addplot[
        mark=o,
        line width=0.75pt,
        only marks,
        point meta=explicit symbolic,
        violet,
        nodes near coords={
            \pgfkeys{/print label/\pgfplotspointmeta/.try}
        },
        every node near coord/.append style={font=\small, anchor=south,yshift=+2pt, xshift=-1pt, inner sep=2}
    ]
        table[header=false,meta index=0,x index=1,y index=2]{
			FCRN 12840000000 4.35
        };
\addplot[
        mark=o,
        line width=0.75pt,
        only marks,
        point meta=explicit symbolic,
        blue,
        nodes near coords={
            \pgfkeys{/print label/\pgfplotspointmeta/.try}
        },
        every node near coord/.append style={font=\small, anchor=south,yshift=1pt}
    ]
        table[header=false,meta index=0,x index=1,y index=2]{
			FCRN15 2011000000 4.24
        };
\addplot[
        mark=o,
        line width=0.75pt,
        only marks,
        point meta=explicit symbolic,
        cyan,
        nodes near coords={
            \pgfkeys{/print label/\pgfplotspointmeta/.try}
        },
        every node near coord/.append style={font=\small, anchor=south,yshift=4.5pt, xshift=4pt}
    ]
        table[header=false,meta index=0,x index=1,y index=2]{
			gGCRN16 583000000 4.39	
        };
\addplot[
        mark=o,
        line width=0.75pt,
        only marks,
        point meta=explicit symbolic,
        magenta,
        nodes near coords={
            \pgfkeys{/print label/\pgfplotspointmeta/.try}
        },
        every node near coord/.append style={font=\small, anchor=south,yshift=3pt, xshift=+13pt}
    ]
        table[header=false,meta index=0,x index=1,y index=2]{
			CRUSE 685000000 4.32
        };
\addplot[
        mark=o,
        line width=0.75pt,
        only marks,
        point meta=explicit symbolic,
        orange,
        nodes near coords={
            \pgfkeys{/print label/\pgfplotspointmeta/.try}
        },
        every node near coord/.append style={font=\small, anchor=south,yshift=-14pt}
    ]
        table[header=false,meta index=0,x index=1,y index=2]{
			FDKF_ 330000000 3.39
        };

\draw [densely dashed, violet] (axis cs:12840000000,3.94) -- (axis cs:12840000000,4.35);
\draw [densely dashed, blue] (axis cs:2011000000,3.80) -- (axis cs:2011000000,4.24);
\draw [densely dashed, cyan] (axis cs:583000000,3.98) -- (axis cs:583000000,4.39); %gG
\draw [densely dashed, magenta] (axis cs:685000000,3.92) -- (axis cs:685000000,4.32);% Cruse
\draw [densely dashed, orange] (axis cs:330000000,3.63) -- (axis cs:330000000,3.39);% Kalman

\end{axis}

\draw (0.125,-0.15) node (t) {\rotatebox[origin=c]{270}{\rule{3mm}{0mm}{$\approx$}}};

\end{tikzpicture}
    % \end{subfigure}%
    \hspace{-1.4em}%
    % \begin{subfigure}
        	\usetikzlibrary{dsp}
	\usetikzlibrary{chains}
	\usetikzlibrary{shapes.geometric}
	\usetikzlibrary{calc}

\pgfkeys{
    /prepare label/.style={
        /print label/\detokenize{#1}/.code={\ttfamily\detokenize{#1}}
    },
    /prepare label/.list={FCRN,FCRN15, gGCRN16, CRUSE, Kalman}
}

\begin{tikzpicture}

\definecolor{color0}{rgb}{0.67843137254902,0.847058823529412,0.901960784313726}

\begin{axis}[
width=.60*\columnwidth,
height=.60*\columnwidth,
legend cell align={left},
legend style={fill opacity=1.0, draw opacity=1, text opacity=1, draw=white!80!black},
log basis x={10},
tick align=outside,
tick pos=left,
xlabel={\#Parameters},
% xmajorgrids,
xmin=700000, xmax=5000000,
xmode=log,
tick style={color=black},
y label style={anchor=north, xshift=17mm, yshift=-3mm},
ylabel={\rotatebox[origin=c]{270}{{MOS}}},
ymajorgrids,
yminorgrids,
minor y tick num=4,
ymin=3.3, ymax=4.7,
extra y ticks={4.6},
extra y tick labels={
    {}
},
extra y tick style={%
    major tick length=3pt,
},
% ytick style={color=black},
extra x ticks={770000},
extra x tick style={%
    every tick/.append style={white, line width=2.7pt},
    yshift=1pt,
    % every tick/.append style={red},
    grid=none,
},
extra x tick labels={
    {}
    % \rotatebox[origin=c]{270}{\rule{3mm}{$\approx$}}
},
]
\addplot[
        mark=star,
        line width=0.75pt,
        only marks,
        point meta=explicit symbolic,
        cyan,
        nodes near coords={
            \pgfkeys{/print label/\pgfplotspointmeta/.try}
        },
        every node near coord/.append style={font=\small, anchor=south,yshift=-18pt, fill=white, inner sep=2}
    ]
        table[header=false,meta index=0,x index=1,y index=2]{
			gGCRN16_ 1300000 3.48	
        };
\addplot[
        mark=star,
        line width=0.75pt,
        only marks,
        point meta=explicit symbolic,
        magenta,
        nodes near coords={
            \pgfkeys{/print label/\pgfplotspointmeta/.try}
        },
        every node near coord/.append style={font=\small, anchor=south,yshift=-10pt, xshift=2pt, fill=white, inner sep=1pt}
    ]
        table[header=false,meta index=0,x index=1,y index=2]{
			CRUSE 1900000 3.48
        };
\addplot[
        mark=star,
        line width=0.75pt,
        only marks,
        point meta=explicit symbolic,
        blue,
        nodes near coords={
            \pgfkeys{/print label/\pgfplotspointmeta/.try}
        },
        every node near coord/.append style={font=\small, anchor=south,yshift=-13pt, fill=white, inner sep=2}
    ]
        table[header=false,meta index=0,x index=1,y index=2]{
			FCRN15_ 1000000 3.49
        };
\addplot[
        mark=star,
        line width=0.75pt,
        only marks,
        point meta=explicit symbolic,
        violet,
        nodes near coords={
            \pgfkeys{/print label/\pgfplotspointmeta/.try}
        },
        every node near coord/.append style={font=\small, anchor=south,yshift=-11pt, fill=white, inner sep = 1pt}
    ]
        table[header=false,meta index=0,x index=1,y index=2]{
			FCRN 3700000 3.49
        };
\addplot[
        mark=star,
        line width=0.75pt,
        only marks,
        point meta=explicit symbolic,
        orange,
        nodes near coords={
            \pgfkeys{/print label/\pgfplotspointmeta/.try}
        },
        every node near coord/.append style={font=\small, anchor=south,yshift=-10pt, xshift=17pt, fill=white, inner sep=1pt}
    ]
        table[header=false,meta index=0,x index=1,y index=2]{
			Kalman 720000 3.48
        };
\end{axis}

\begin{axis}[
width=.60*\columnwidth,
height=.60*\columnwidth,
  axis y line*=right,
  axis x line=none,
xmin=700000, xmax=5000000,
xmode=log,
  ymin=3.3, ymax=4.7,
  y axis line style=black,
  every tick label/.append style={ticks=none},
  legend pos=north east
]
\addlegendimage{only marks, mark=o, black} \addlegendentry{ST\,Echo}
\addlegendimage{only marks, mark=star, black}\addlegendentry{ST\,Other}
% \addplot[only marks, mark=o,black, point meta=explicit symbolic,
% 		nodes near coords={
%        		\pgfkeys{/print label/\pgfplotspointmeta/.try}
%   		},
%         every node near coord/.append style={font=\small, anchor=south,yshift=-13pt}]
%         table[header=false,meta index=0,x index=1,y index=2]{
% 			CRUSE5 183307688 14.21
%         }; %\addlegendentry{\(\displaystyle \Delta\)SNR}
\addplot[
        mark=o,
        line width=0.75pt,
        only marks,
        point meta=explicit symbolic,
        violet,
        nodes near coords={
            \pgfkeys{/print label/\pgfplotspointmeta/.try}
        },
        every node near coord/.append style={font=\small, anchor=south,yshift=-5pt, xshift=-22pt, fill=white, inner sep=2}
    ]
        table[header=false,meta index=0,x index=1,y index=2]{
			FCRN_ 3700000 4.21
        };
\addplot[
        mark=o,
        line width=0.75pt,
        only marks,
        point meta=explicit symbolic,
        blue,
        nodes near coords={
            \pgfkeys{/print label/\pgfplotspointmeta/.try}
        },
        every node near coord/.append style={font=\small, anchor=south,yshift=2.5pt, xshift=0pt, fill=white, inner sep = 0.5pt}
    ]
        table[header=false,meta index=0,x index=1,y index=2]{
			FCRN15 1000000 4.28
        };
\addplot[
        mark=o,
        line width=0.75pt,
        only marks,
        point meta=explicit symbolic,
        magenta,
        nodes near coords={
            \pgfkeys{/print label/\pgfplotspointmeta/.try}
        },
        every node near coord/.append style={font=\small, anchor=south,yshift=0pt, xshift=11pt}
    ]
        table[header=false,meta index=0,x index=1,y index=2]{
			CRUSE_ 1900000 4.09
        };
\addplot[
        mark=o,
        line width=0.75pt,
        only marks,
        point meta=explicit symbolic,
        cyan,
        nodes near coords={
            \pgfkeys{/print label/\pgfplotspointmeta/.try}
        },
        every node near coord/.append style={font=\small, anchor=south, yshift=3pt, xshift=8pt, fill=white, inner sep = 0.5pt}
    ]
        table[header=false,meta index=0,x index=1,y index=2]{
			gGCRN16 1300000 4.08	
        };
\addplot[
        mark=o,
        line width=0.75pt,
        only marks,
        point meta=explicit symbolic,
        orange,
        nodes near coords={
            \pgfkeys{/print label/\pgfplotspointmeta/.try}
        },
        every node near coord/.append style={font=\small, anchor=south,xshift=12pt}
    ]
        table[header=false,meta index=0,x index=1,y index=2]{
			Kalman_ 720000 4.03
        };
\draw [densely dashed, magenta] (axis cs:1900000,3.48) -- (axis cs:1900000,4.09);% Cruse
\draw [densely dashed, cyan] (axis cs:1300000,3.48) -- (axis cs:1300000,4.08); %gG
\draw [densely dashed, blue] (axis cs:1000000,3.49) -- (axis cs:1000000,4.28);
\draw [densely dashed, violet] (axis cs:3700000,3.49) -- (axis cs:3700000,4.21);
\draw [densely dashed, orange] (axis cs:720000,3.48) -- (axis cs:720000,4.03);
\end{axis}

\draw (0.17,-0.15) node (t) {\rotatebox[origin=c]{270}{\rule{3mm}{0mm}{$\approx$}}};

\end{tikzpicture}
        % \includegraphics[width=.49\columnwidth]{fig/complexity_plot_ST}
    % \end{subfigure}%
    \vspace{-4mm}
    \caption{Model performance for DT (left) and ST (right) AECMOS metrics over \#FLOPS and \#parameters, evaluated on the \textbf{test set} $\mathcal{D}_\mathrm{test}$.}
    \label{fig:complexity_plot}
    \vspace{-4.5mm}
\end{figure}
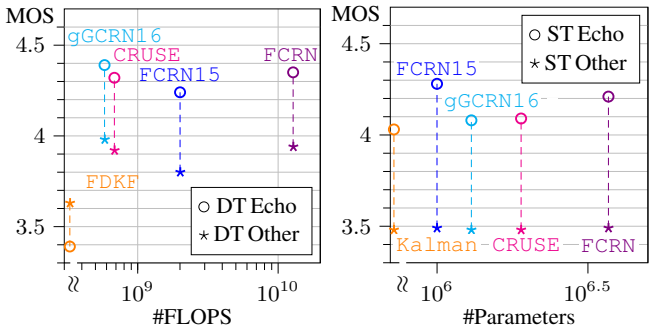

Fig.\,\ref{fig:complexity_plot} shows various tested model architectures evaluated by the AECMOS metrics. As before, we only compare models trained on the same MCS (16/0/0). We include the frequency domain Kalman filter (FDKF) after \cite{Franzen2018a} without their postfilter as a classical AEC system of negligible computational complexity and model size.

We can see that compared to the baseline {\tt FCRN}, the {\tt gGCRN16} architecture greatly reduces the parameter count (65\% less) and FLOPS (95\% less) while delivering high DT performance. While ST\,Echo degrades a bit, the STFE echo suppression still stays on a high level where the remaining residual echo should be easily removable by a postfilter. {\it We again outperform or are on par with the {\tt CRUSE} architecture in all metrics.} As with the dev set, all models show next-to-perfect performance on the STNE condition with only marginal deviations on $\mathcal{D}_\mathrm{test}$ as well. All DAEC models considerably outperform the FDKF in DT.

Table \ref{tab:results_tune_test} shows the exemplary impact of two fine-tunings with condition-aware component losses after (\ref{eq:cda_loss}). All shown models are fine-tuned, either using the standard training loss ($\alpha_b = \beta_b = 0$) or context-aware losses (marked by the identifier CA). The specific weight values for context aware-losses are defined as follows: For (15/1/0), we choose $\alpha_\mathrm{DT}=0.2$, $\beta_{\mathrm{DT}}=0.2$, $\alpha_{\mathrm{STFE}}=0.2$, $\beta_{\mathrm{STFE}}=0$, for (16/0/0) we choose $\alpha_{\mathrm{DT}}=0.33$, $\beta_{\mathrm{DT}}=0.0$. 

\begin{table}[t]
	\setlength{\tabcolsep}{.35em}
    \vspace{-3.5mm}
	\caption{Evaluation of fine-tuning {\tt gGCRN16} on two MCSs, using either the standard training loss defined in Section \ref{ssec:training} or condition-aware component losses (CA) defined in (\ref{eq:cda_loss}) on the \textbf{test set} $\mathcal{D}_{\mathrm{test}}$. Best results are {bold}, second best {underlined}.}
	\vspace{0.5mm}
    \centering
	\newcolumntype{R}{>{\raggedleft\arraybackslash}X}
\newcolumntype{C}{>{\center\arraybackslash}X}
\newcommand{\bftab}{\fontseries{b}\selectfont}

\begin{tabular}{l ccccc} 

	\toprule
\multirow{2}{*}{\vtop{\hbox{\strut{\textbf{Model,}}} \hbox{\strut{\textbf{MCS}}}}} & \multicolumn{5}{c}{\textbf{Double-Talk}}\\

\cmidrule(lr){2-6}

& PESQ & PESQ$_\mathrm{BB}$ & ERLE$_\mathrm{BB}$ & DT Other & DT Echo\\
\cmidrule(lr){1-6}

 15/1/0, CA & \underline{2.33}          & 3.51              & \textbf{17.56}    & \underline{3.97}  & \textbf{4.45} \\
 15/1/0        & \textbf{2.36} & 3.68              & \underline{15.76} & 3.95              & 4.37 \\
 16/0/0        & \textbf{2.36} & \underline{3.69}  & 15.58             & \textbf{3.98}     & \underline{4.40}\\
 16/0/0, CA & \textbf{2.36} & \textbf{3.78}     & 14.71             & 3.95              & 4.36\\
 
\bottomrule
\end{tabular}
    \vspace{-2.5mm}
	\label{tab:results_tune_test}
\end{table}

We see that the main distinction of the models' behaviour can be observed in the component metrics PESQ$_\mathrm{BB}$ and ERLE$_\mathrm{BB}$. The fine-tuned model of (15/1/0) leans more towards aggressive echo suppression, while the fine-tuned (16/0/0) model preserves more NE speech at the cost of a weaker AES. This trade-off control can be leveraged to adjust the model performance towards a specific task. For example, we could tune a model to remove just as much echo as necessary for a postfilter to reliably suppress residual echo without sacrificing NE speech too much.

% \begin{table}[t]
% 	\setlength{\tabcolsep}{.35em}
% 	\caption{Files per batch drawn from the respective condition for trained gGCRN16 variations. Model 1 draws conditions evenly from each condition at random.}
% 	\vspace{0.5mm}
%     \centering
% 	\input{tables/TABLE_distDEF.tex}
%     \vspace{-3mm}
% 	\label{tab:distDef}
% \end{table}

% \begin{table}[t]
% 	\setlength{\tabcolsep}{.35em}
% 	\caption{Components loss weights used for gGCRN16 fine-tuning.}
% 	\vspace{0.5mm}
%     \centering
% 	\input{tables/TABLE_weights.tex}
%     \vspace{-3mm}
% 	\label{tab:weights}
% \end{table}

% \begin{table}[b]
% 	\setlength{\tabcolsep}{.35em}
% 	\caption{Research: on $\mathcal{D}_{\mathrm{test}}$}
% 	\vspace{0.5mm}
% 	\input{tables/TABLE_dist_test.tex}
%     \vspace{-3mm}
% 	\label{tab:TABLE_dist_test}
% \end{table}

\section{Conclusions}
\label{sec:results}

This work demonstrates the importance of a high DT proportion in the DAEC training data.
We further present modifications to the {\tt FCRN15} architecture that not only reduce computational complexity by more than 70\%, but also improve DT performance. The final model also outperforms the {\tt CRUSE DAEC-64} model, although it has 30\% less parameters and 15\% less computational complexity. 
The proposed condition-aware loss allows further control on the trade-off between NE preservation and echo suppression.

\newpage

% \begin{equation}
%   \label{eqn:wave_equation}
%     \Delta^2p(x,y,z,t)-
%     \displaystyle\frac{1}{c^2}\frac{\partial^2p(x,y,z,t)}{\partial t^2}=0,
% \end{equation}

% -------------------------------------------------------------------------
\bibliographystyle{IEEEtran}
\bibliography{refs, ifn_spaml_bibliography}

\begin{thebibliography}{10}
\providecommand{\url}[1]{#1}
\def\UrlFont{\rmfamily}
\providecommand{\newblock}{\relax}
\providecommand{\bibinfo}[2]{#2}
\providecommand\BIBentrySTDinterwordspacing{\spaceskip=0pt\relax}
\providecommand\BIBentryALTinterwordstretchfactor{4}
\providecommand\BIBentryALTinterwordspacing{\spaceskip=\fontdimen2\font plus
\BIBentryALTinterwordstretchfactor\fontdimen3\font minus
  \fontdimen4\font\relax}
\providecommand\BIBforeignlanguage[2]{{%
\expandafter\ifx\csname l@#1\endcsname\relax
\typeout{** WARNING: IEEEtran.bst: No hyphenation pattern has been}%
\typeout{** loaded for the language `#1'. Using the pattern for}%
\typeout{** the default language instead.}%
\else
\language=\csname l@#1\endcsname
\fi
#2}}

\bibitem{SpeexDSP}
J.-S. Soo and K.~Pang, ``{Multidelay Block Frequency Domain Adaptive Filter},''
  \emph{IEEE Transactions on Acoustics, Speech, and Signal Processing},
  vol.~38, no.~2, pp. 373--376, Feb. 1990.

\bibitem{enzner_vary_fdaf}
G.~Enzner and P.~Vary, ``{Frequency-Domain Adaptive Kalman Filter for Acoustic
  Echo Control in Hands-Free Telephones},'' \emph{{S}ignal {P}rocessing},
  vol.~86, no.~6, pp. 1140--1156, June 2006.

\bibitem{Valin_AEC}
J.-M. Valin, S.~Tenneti, K.~Helwani, U.~Isik, and A.~Krishnaswamy,
  ``{Low-Complexity, Real-Time Joint Neural Echo Control and Speech Enhancement
  Based On Percepnet},'' in \emph{Proc. of ICASSP}, Toronto, Canada, June 2021,
  pp. 7133--7137.

\bibitem{Zhang2022a}
S.~Zhang, Z.~Wang, J.~Sun, Y.~Fu, B.~Tian, Q.~Fu, and L.~Xie, ``{Multi-Task
  Deep Residual Echo Suppression with Echo-Aware Loss},'' in \emph{Proc. of
  ICASSP}, Singapore, May 2022, pp. 9127--9131.

\bibitem{Revach2021}
G.~Revach, N.~Shlezinger, R.~J.~G. van Sloun, and Y.~C. Eldar, ``{Kalmannet:
  Data-Driven Kalman Filtering},'' in \emph{Proc. of ICASSP}, Toronto, Canada,
  June 2021, pp. 3905--3909.

\bibitem{westhausen21_icassp}
N.~L. Westhausen and B.~T. Meyer, ``{Acoustic Echo Cancellation with the
  Dual-Signal Transformation LSTM Network},'' in \emph{Proc. of ICASSP},
  Toronto, Canada, June 2021, pp. 7138--7142.

\bibitem{seidel21_interspeech}
E.~Seidel, J.~Franzen, M.~Strake, and T.~Fingscheidt, ``{Y$^2$-Net FCRN for
  Acoustic Echo and Noise Suppression},'' in \emph{Proc. of Interspeech}, Brno,
  Czech Republic, Oct. 2021, pp. 4763--4767.

\bibitem{Braun2022}
S.~Braun and M.~L. Valero, ``{Task Splitting for DNN-based Acoustic Echo and
  Noise Removal},'' in \emph{Proc. of IWAENC}, Bamberg, Germany, Sept. 2022,
  pp. 1--5.

\bibitem{Wang2022}
H.~Zhang and D.~Wang, ``{Neural Cascade Architecture for Joint Acoustic Echo
  and Noise Suppression},'' in \emph{Proc. of ICASSP}, Singapore, May 2022, pp.
  671--675.

\bibitem{Strake2020}
M.~Strake, B.~Defraene, K.~Fluyt, W.~Tirry, and T.~Fing\-scheidt, ``{Fully
  Convolutional Recurrent Networks for Speech Enhancement},'' in \emph{Proc.\
  of ICASSP}, Barcelona, Spain, May 2020, pp. 6674--6678.

\bibitem{cutler2022AEC}
R.~Cutler, A.~Saabas, T.~Parnamaa, M.~Purin, H.~Gamper, S.~Braun, K.~Sorensen,
  and R.~Aichner, ``{ICASSP 2022 Acoustic Echo Cancellation Challenge},'' in
  \emph{Proc. of ICASSP}, Singapore, May 2022, pp. 9107--9111.

\bibitem{pfeifenberger21_interspeech}
L.~Pfeifenberger, M.~Zoehrer, and F.~Pernkopf, ``{Acoustic Echo Cancellation
  with Cross-Domain Learning},'' Brno, Czech Republic, Oct. 2021, pp.
  4753--4757.

\bibitem{Strake2022}
M.~Strake, A.~Behlke, and T.~Fingscheidt, ``{Self-Attention With Restricted
  Time Context And Resolution In Dnn Speech Enhancement},'' in \emph{Proc. of
  IWAENC}, Bamberg, Germany, Sept. 2022, pp. 1--5.

\bibitem{Strake2020a}
M.~Strake, B.~Defraene, K.~Fluyt, W.~Tirry, and T.~Fing\-scheidt,
  ``{INTERSPEECH 2020 Deep Noise Suppression Challenge: A Fully Convolutional
  Recurrent Network (FCRN) for Joint Dereverberation and Denoising},''
  Shanghai, China, Oct. 2020, pp. 2467--2471.

\bibitem{Tan2020}
K.~Tan and D.~Wang, ``{Learning Complex Spectral Mapping With Gated
  Convolutional Recurrent Networks for Monaural Speech Enhancement},''
  \emph{IEEE Transactions on Audio, Speech and Language Processing}, vol.~28,
  pp. 380--390, Nov. 2020.

\bibitem{VCTK}
J.~Yamagishi, C.~Veaux, and K.~MacDonald, ``{CSTR VCTK Corpus: English
  Multi-speaker Corpus for CSTR Voice Cloning Toolkit},'' {University of
  Edinburgh. The Centre for Speech Technology Research}, 2017.

\bibitem{Zhang2020c}
H.~Zhang and D.~Wang, ``{A Deep Learning Approach to Active Noise Control},''
  in \emph{Proc. of Interspeech}, Shanghai, China, Oct. 2020, pp. 1141--1145.

\bibitem{Klippel2006}
W.~Klippel, ``{Tutorial: Loudspeaker Nonlinearities - Causes, Parameters,
  Symptoms},'' \emph{Journal of the Audio Engineering Society}, vol.~54,
  no.~10, pp. 907--939, Oct. 2006.

\bibitem{image_method}
J.~B. Allen and D.~A. Berkley, ``{Image Method for Efficiently Simulating
  Small-Room Acoustics},'' \emph{{J.\ Acoust.\ Soc.\ Am.}}, vol.~65, no.~4, pp.
  943--950, 1979.

\bibitem{Thiemann2013}
J.~Thiemann, N.~Ito, and E.~Vincent, ``{The Diverse Environments Multi-Channel
  Acoustic Noise Database: A Database of Multichannel Environmental Noise
  Recordings},'' \emph{J.\ Acoust.\ Soc.\ Am.}, vol. 133, no.~5, pp.
  3591--3591, 2013.

\bibitem{QUT}
D.~B. Dean, S.~Sridharan, R.~J. Vogt, and M.~W. Mason, ``{The QUT-NOISE-TIMIT
  Corpus for the Evaluation of Voice Activity Detection Algorithms},''
  Makuhari, Japan, Sept. 2010, p. 3110–3113.

\bibitem{ETSI2008}
\emph{{Speech Processing, Transmission and Quality Aspects (STQ); Speech
  Quality Performance in the Presence of Background Noise; Part 1: Background
  Noise Simulation Technique and Background Noise Database}}, ETSI EG 202
  396-1, Sept. 2008.

\bibitem{TIMIT}
J.~S. Garofolo, L.~F. Lamel, W.~M. Fisher, J.~G. Fiscus, and D.~S. Pallett,
  ``{TIMIT Acoustic-Phonetic Continous Speech Corpus},'' {Linguistic Data
  Consortium, Philadelphia, PA, USA}, 1993.

\bibitem{Jung2013}
M.-A. Jung and L.~R.~T. Fingscheidt, ``{Towards Reproducible Evaluation of
  Automotive Hands-Free Systems in Dynamic Conditions},'' in \emph{Proc. of
  ICASSP}, Vancouver, Canada, May 2013, pp. 8144--8148.

\bibitem{Zoellzer2003}
U.~Zöllzer, \emph{{DAFX: Digital Audio Effects}}.\hskip 1em plus 0.5em minus
  0.4em\relax Wiley, 2003.

\bibitem{ITU_P862.2_Corr}
``{ITU-T Rec. P.862.2 Corrigendum 1, Wideband Extension to Rec. P.862 for the
  Assessment of Wideband Telephone Networks and Speech Codecs},'' ITU-T, Oct.
  2017.

\bibitem{Vary2006}
P.~Vary and R.~Martin, \emph{Digital Speech Transmission}.\hskip 1em plus 0.5em
  minus 0.4em\relax John Wiley \& Sons, Ltd, 2006.

\bibitem{purin2021aecmos}
M.~Purin, S.~Sootla, M.~Sponza, A.~Saabas, and R.~Cutler, ``{AECMOS: A Speech
  Quality Assessment Metric for Echo Impairment},'' in \emph{Proc. of ICASSP},
  Singapore, May 2022, pp. 901--905.

\bibitem{fingscheidt_signalseparation}
T.~Fingscheidt and S.~Suhadi, ``{Quality Assessment of Speech Enhancement
  Systems by Separation of Enhanced Speech, Noise, and Echo},'' {Antwerp,
  Belgium}, Aug. 2007, pp. 818--821.

\bibitem{adam_optimizer}
D.~P. Kingma and J.~Ba, ``{Adam: A Method for Stochastic Optimization},'' in
  \emph{Proc. of ICLR}, San Diego, USA, May 2015, pp. 1--15.

\bibitem{Abadi2016}
M.~Abadi \emph{et~al.}, ``{TensorFlow: Large-Scale Machine Learning on
  Heterogeneous Distributed Systems},'' \emph{arXiv}, Mar. 2016.

\bibitem{Franzen2018a}
J.~Franzen and T.~Fing\-scheidt, ``{An Efficient Residual Echo Supression for
  Multi-Channel Acoustic Echo Cancellation Based on the Frequency-Domain
  Adaptive Kalman Filter},'' in \emph{Proc. of ICASSP}, Calgary, AB, Canada,
  Apr. 2018, pp. 226--230.

\end{thebibliography}

\end{sloppy}
\end{document}